\documentclass[conference]{IEEEtran}
\ifCLASSOPTIONcompsoc
  \usepackage[nocompress]{cite}
\else
  \usepackage{cite}
\fi

\usepackage{times}
\usepackage{graphicx}
\usepackage{subfigure}
\usepackage{epstopdf}		
\usepackage[cmex10]{amsmath}
\usepackage{relsize}
\usepackage{longtable}
\usepackage{pdfpages}
\usepackage{bbding}
\usepackage{pifont}
\usepackage{wasysym}
\usepackage{amssymb}
\usepackage{listings}
\usepackage{url}
\usepackage{color}
\usepackage{setspace}
\usepackage{float}
\usepackage{multirow}
\usepackage{enumerate}

\newcommand\tab[1][1cm]{\hspace*{#1}}

\begin{document}
%
\title{
Learning-based Dynamic Cache Management in a Cloud
}



\author{Jinhwan Choi\IEEEauthorrefmark{1}\, Yu Gu\IEEEauthorrefmark{2}\, Jinoh Kim\IEEEauthorrefmark{1} \\
\IEEEauthorblockA{\\Texas A\&M University, Commerce, TX 75428, USA \IEEEauthorrefmark{1}\\
VISA Inc., Austin, USA \IEEEauthorrefmark{2} \\
 Email: jchoi8@leomail.tamuc.edu, yugu1@visa.com, jinoh.kim@tamuc.edu}}

\maketitle

\begin{abstract}
Caches are an important component of modern computing systems given their significant impact on performance. In particular, caches play a key role in the cloud due to the nature of large-scale, data-intensive processing. One of the key challenges for the cloud providers is how to share the caching capacity among tenants, under the circumstance that each often requires a different degree of quality of service (QoS) with respect to data access performance. The invariant is that the individual tenants' QoS requirements should be satisfied while the cache usage is optimized in a system-wide manner.
In this paper, we introduce a learning-based approach for dynamic cache management in a cloud, which is based on the estimation of  data access pattern of a tenant and the prediction of  cache performance for the access pattern in question. We consider a variety of probability distributions to estimate the data access pattern, and examine a set of learning-based regression techniques to predict the cache hit rate for the access pattern.
The predicted cache hit rate is then used to make a decision whether reallocating cache space is needed  to meet the QoS requirement for the tenant.
Our experimental results with an extensive set of synthetic traces and the YCSB benchmark show that the proposed method consistently optimizes the cache space while satisfying the QoS requirement.
\end{abstract}


%

%
\IEEEpeerreviewmaketitle

\section{Introduction}
	\label{sec:intro}

Caches are an important component of modern computing systems given their significant impact on the data access performance, which is more critical in a large-scale data processing environment. 
For example, leading Web service providers such as Google, Facebook and Amazon largely rely on in-memory processing using  Memcached~\cite{Memcached,ElMem} and Redis~\cite{Redis}.
Even small-scale providers employ the caching services through Amazon ElastiCache~\cite{ElastiCache}, Memcachier~\cite{MemCachier}, and so forth. 
This is because that the efficient use of in-memory caching has a significant impact on the performance as the reading from the storage is much more expensive than from the cache. 
In addition to the performance benefits, using a cache reduces the load in the back-end data servers giving greater scalability.
Since the cloud often requires processing massive data (e.g., for hosting Web or e-Commerce services), 
in-memory caching is an essential part to improve the data access performance. 

As the number of users and network traffic increases, service providers may need to expand the infrastructure. 
However, adding hardware resources infinitely for better performance is costly and impossible.  
Thus, it is important to use the limited resources more in an efficient way.
In that sense, cache management plays a key role to better utilize the limited cache spaces, in order to maximize the performance and efficiency of data access. 
Cache management consists of a set of functions including  cache space (re)allocation and eviction. 
In this work, we focus on the problem of cache space optimization under the assumption of the use of in-memory caching in a cloud. 

A crucial concern for cache management in a cloud is that the caches are shared by multiple tenants 
that may have different
quality of service (QoS) requirements with respect to data access performance.
Thus, it is necessary to meet the per-tenant requirement in addition to the optimal use of the cache in a system-wide manner.
A simple technique to deal with this problem is {\em static caching} that allocates the exclusively dedicated cache blocks to tenants based on the individual QoS specifications.
However, it may not be straightforward to determine how many cache blocks would be needed for a tenant without having the knowledge of data access pattern. 
Moreover, the access pattern can be changed over time, and the initially assigned cache blocks can be too small violating the performance requirement or too large causing the waste of the limited system resource. 
{\em Dynamic caching} (re)allocates cache space based on the demand of tenants dynamically, and it would be beneficial
not only to meet the per-tenant requirement but also to improve the overall system performance.
Hence, it is essential to properly capture the demand of cache space when using dynamic caching.
Another alternative is to use a global cache shared among tenants but isolation guarantee, which is one of the desirable properties in a cloud, cannot be provided~\cite{FairRide,OPUS}. 
In this work, we take dynamic caching into account to deal with the problem of cache management with the benefit of low overhead in the cache replacement time since the interval of cache re-allocation is relatively infrequent and does not need to be performed at every eviction. 

\begin{figure}[!tb]
\centering
\includegraphics[width=1.0\columnwidth]{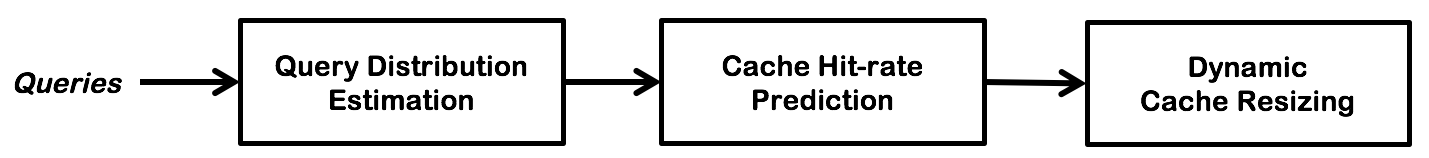}
\caption{
Process model for dynamic cache management
}
\label{fig:process_model}
\end{figure}

For dynamic caching in a cloud,  
there has been a body of work in the past few years, 
mainly by estimating the cache performance of individual tenants (or applications)~\cite{Dynacache,FairRide,ChocklerLV11,Chockler:2010}, 
However, previous studies are largely limited with the lack of well-defined models for the estimation of data access patterns and the prediction of cache performance for a given access pattern. 
With the recent disruptive advance of machine learning technologies, 
we take a learning-based approach to cache management in a cloud. 
%
To estimate data access patterns varied over time, we consider a set of probabilistic distributions (uniform, Gaussian, exponential, and Zipf). 
The information of the estimated access pattern is then used to predict the cache size to meet the specified data access performance, 
and we assume the QoS is defined with the required cache hit rate.
For this purpose, we examine various regression methods,
including Support Vector Regression (SVR), Gaussian Process Regression (GPR), and Fully Connected Neural Network (FCN).   
Figure~\ref{fig:process_model} shows our  process model to implement dynamic caching in a cloud with the aforementioned functions of access pattern estimation and cache performance prediction. 
For a query-based transaction system, we assume that the access pattern is inferred based on the query key distribution.
From the estimated distribution, cache hit rate is predicted to determine the right size of cache space to meet the given QoS requirement.
Finally, cache resizing takes place if needed. 

The key contributions of this paper are summarized as follows:

\begin{itemize}
   \item We formulate a problem of dynamic caching in a cloud, with two objectives of QoS guarantee and optimization of the cache space in the system;
   \item We present a method for the query distribution estimation with  a set of probabilistic  distributions including normal, Gaussian, exponential, and Zipf. 
	We demonstrate that our estimation method using the Kolmogorov-Smirnov Test (KS-Test)~\cite{Choi:LBNL:2013} can yield accurate estimations even with a small number of query samples ($\le 1000$ samples), which is beneficial for responding to the temporal pattern changes in a timely manner;
   \item We examine a set of learning-based regression techniques including SVR, GPR, and FCN, to predict the cache hit rate based on the estimated query distribution.
	Our evaluation supports the effectiveness of the FCN model across the diverse distributions with respect to the prediction performance; 
   \item We evaluate our proposed process for dynamic caching with an extensive set of synthetic traces and Yahoo! Cloud Serving Benchmark (YCSB)~\cite{YCSB}, on a cluster computing system with Open Source Apache Ignite~\cite{ApacheIgnite} as the in-memory caching infrastructure.
	The evaluation results confirm that the proposed method consistently optimizes the cache space, maintaining  the required cache hit rate as the QoS requirement. 
	
\end{itemize}

This paper is organized as follows. 
In the next section, we specify the description of the problem tackled in this study and provide a summary of the closely related studies in the area of cache management in the cloud. 
Section~\ref{sec:design} presents our proposed process model with the details of the functional elements.
We report the evaluation results in Section~\ref{sec:eval} conducted with the synthetic traces and YCSB benchmark tools.
Finally, we conclude our presentation in Section~\ref{sec:conc} with a summary of the work and  future directions.

\section{Background}
	\label{sec:bg}

\subsection{Problem Statement}
	\label{sec:prob}
	
In this work, we assume a cloud system equipped with a large-scale in-memory cache. 
A tenant loads their data to the cloud and accesses them. 
We assume a read-many, write-rare environment through key-value stores like a database transaction system.
Hence, the cache performance is a predominant factor for the overall system performance.
The tenant is associated with an agreement (SLA) that specifies a set of QoS requirements including data access performance.
In this study, we focus on the cache hit rate to measure the data access performance.

\begin{table}[!tb]
\caption{Notation}
\label{tab:notation}
\centering
\begin{tabular}{l|l}
\hline
Notation & Description \\
\hline
$T_k$ & Tenant $k$ in the tenant set $T$ ($T_k \in T$) \\
$C$ & Total cache space in the system \\
$C_{T_k}$ & Cache size allocated to $T_k$ \\
$h_{T_k}$ & Measured cache hit rate for $T_k$ \\
$H_{T_k}$ & Minimum cache hit rate requirement for $T_k$ \\
$H(x,y)$ & Predicted cache hit rate under cache size $x$ \\
	& and access distribution $y$ \\
$\delta$ & Safety margin \\ 
\hline
\end{tabular}
\end{table}

To formulate, we use the notation defined in Table~\ref{tab:notation}.
A set of tenants $T = \{T_1, T_2, ..., T_n\}$ exist in the cloud with the total cache space of $C$.
Initially, a cache space $C_{T_k}$ is allocated to tenant $T_k$, which can be adjusted over time to meet the specified cache hit rate requirement ($H_{T_k}$). 
The current cache hit rate  ($h_{T_k}$) is periodically measured, and our objective is to keep $h_{T_k} \ge H_{T_k}$ for the tenant. 

It would be easy to guarantee $H_{T_k}$ by simply allocating a plenty of cache space to the tenant which would be much greater than the actual need.
However, there will be a significant waste of the expensive cache resource in that case.
In order to perform the system-wide optimization as well as the minimum QoS guarantee for individual tenants, this work addresses the following optimization problem:

\vspace{.2in}
{\noindent
\tab Minimize \tab $\mathlarger{\sum_{T_k \in T}  C_{T_k}} $ \\
\tab such that \tab $\mathlarger{\sum_{T_k \in T} C_{T_k}  \le C}$ \\
\tab and \tab\tab $\mathlarger{h_{T_k} \ge H_{T_k},  \forall T_k}$ \\ 
}
\vspace{.1in}

If the system has insufficient free cache space to allocate to a certain tenant,
we assume that it is reported to the administrator who conducts the defined policy to deal with such an exceptional situation. 


\subsection{Related Work}
	\label{sec:related}
	
\begin{figure}[!tb]
\centering
\includegraphics[width=.8\columnwidth]{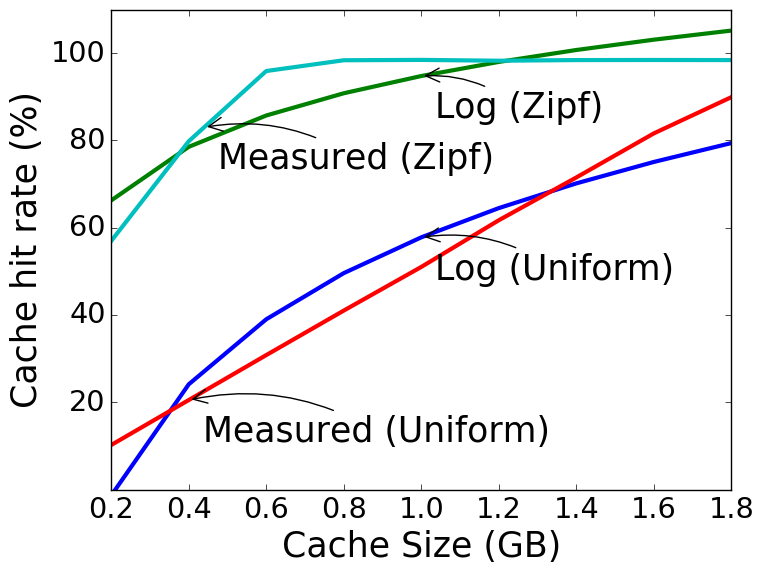}
\caption{
Cache performance estimation using a log function-based least-square fit:
This concave function may not be adequate for cache performance prediction with large errors over 10\% gaps to the measured result. 
}
\label{fig:log_fit}
\end{figure}

There exist several interesting studies for in-memory caching in the cloud environment. This section summarizes the most relevant studies to this research with the differences from our work.

Blaze~\cite{Chockler:2010} is the first generation work for the data caching in a cloud with the consideration of multiple tenants.  
The main objective of this work is not only to guarantee the minimum QoS requirement for each tenant but also to maximize the overall cache hit rate.
To meet the tenant's QoS goal, the authors proposed to estimate the cache hit rate using a log function ($y = a + b\log x$) based on a least-square fit, where the output ($y$) is the predicted cache hit rate from the given cache size ($x$).
In this scheme, determining the parameters ($a$ and $b$) is a non-trivial challenge.
More critically, we observed considerable errors from our preliminary experiment, as shown in Figure~\ref{fig:log_fit} that compares the measured hit rates and the log-based fitted result for two different access patterns. 
The figure shows over 10\% errors for certain predictions. 
As will be demonstrated in Section~\ref{sec:design}, it would be hard to fit the hit rates using a simple concave function, and we employ learning-based techniques to provide greater accuracy and flexibility. 

There have been some studies utilizing the concept of stack distance~\cite{Almasi:2002} to estimate the data access pattern for the cloud cache management. 
In the SC2 work~\cite{ChocklerLV11}, a cache space utility model is introduced to maximize the overall cache hit rate in a multi-tenant environment.
The cache space utility model is a cumulative distribution function of the stack distance hit histogram.
The cumulative curve is then used to figure out the minimum cache space that satisfies the required minimum cache hit rate.
A critical problem referencing to stack distances to estimate data access patterns is the heavy complexity  to keep shadow eviction queues for incoming requests, 
in order to track the location of the request to calculate the stack distance.

Dynacache~\cite{Dynacache} has been proposed to improve the cache hit rates in a multi-application environment.
This technique also relies on stack distances to infer the cache hit rate curve,
and interestingly, it uses an approximation technique using  buckets to reduce the overhead for calculating stack distances.
However, the estimation is still expensive when assuming hundreds of applications running in a system.
Dynacache  defines a metric to identify  applications that can be more benefited from the cache management and stack distances are estimated only for a small set of the identified applications.
While beneficial to reference to the stack distance information, it is heavy to calculate and even the approximation is still expensive. 
We do not rely on stack distances, but employ a set of probability models to estimate tenants' access patterns, with a distribution comparison function. 
This past work also assumes a simple concave function to predict cache hit rates from the given cache size.

A recent work FairRide~\cite{FairRide} tackles a cheating problem that may lead to monopolizing the cache resource by a greedy user.
The cheating problem can happen in an environment that the users often access the same data (e.g., utility programs).
For example, a greedy user can get a free ride by accessing the shared files already in the cache space loaded by other users.
The authors proposed a simple idea to mitigate this problem by imposing a penalty (e.g., delay) to such users.
The work in~\cite{OPUS} further studies on the problem of fair cache allocation for in-memory analytics to prevent from free-rideing manipulations that may result in poor cache utilization. 
In this work, we do not deal with a cheating problem, but it may be an interesting topic for future exploration.


\section{Proposed Design}
	\label{sec:design}

\begin{figure}[!tb]
\centering
\includegraphics[width=1.0\columnwidth]{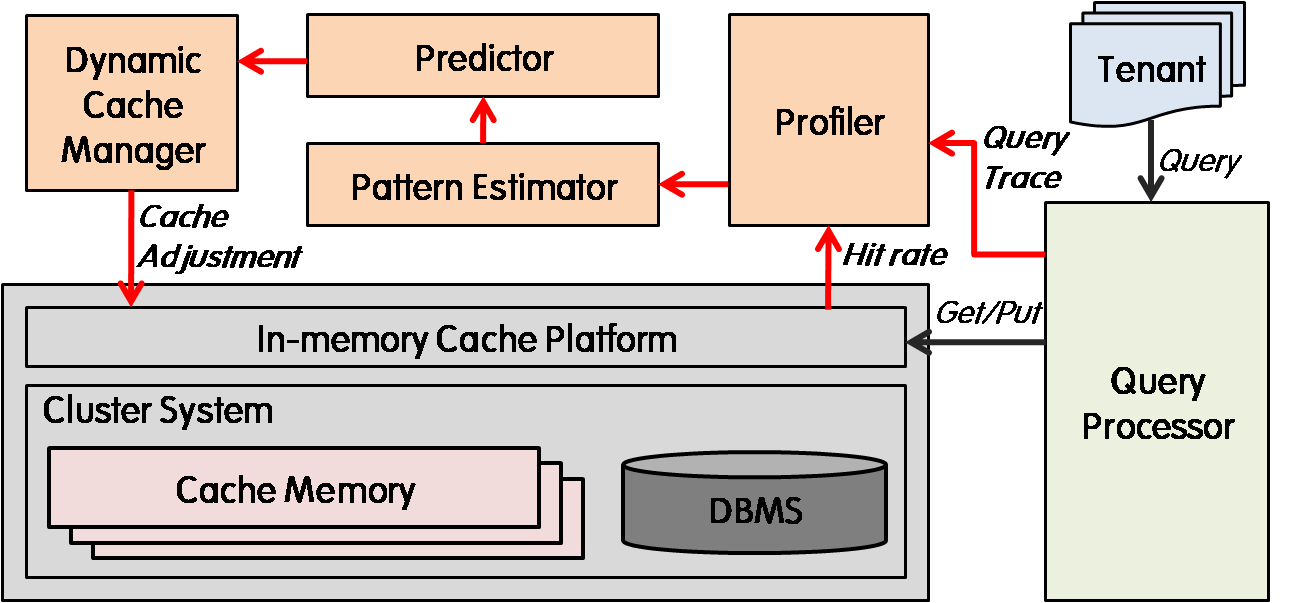}
\caption{
Architecture for dynamic cache management:
{\em Pattern Estimator} estimates the distribution of query keys sampled by {\em Profiler} for each tenant periodically.
Once an estimated distribution information is available, {\em Predictor} makes a prediction using a regression model to calculate  the tenant's cache hit rate with the currently allocated cache size and 
the estimated probability distribution.
{\em Cache Manager} determines whether the tenant's cache needs to be resized or not, based on the prediction result.
}
\label{fig:workflow}
\end{figure}

In this section, we present our design to tackle the problem stated in the previous section. 
Figure~\ref{fig:workflow} shows the proposed architecture for our learning-based dynamic cache management in a cloud. 
The overall scenario to meet the QoS requirement (i.e., cache hit rate) for each tenant is as follows.
{\em Pattern Estimator} estimates the distribution of query keys sampled by {\em Profiler} for each tenant periodically (e.g., at every $k$ number of samples). 
Once the estimation result is available, {\em Predictor} makes a prediction using a regression function to calculate  the tenant's cache hit rate with the currently allocated cache size $C_{T_k}$ and the estimated probability distribution $d$.
In Table~\ref{tab:notation}, we define the predicted cache hit rate for tenant $k$ as $H(C_{T_k}, d)$.
Then the predicted hit rate is compared to the minimal requirement ($H_{T_k}$).
Based on the comparison result, {\em Cache Manager} determines whether the tenant's cache needs to be resized or not.
We next discuss the functional components in the proposed model in detail.

\subsection{Query Distribution Estimation}

\begin{table}[!tb]
\caption{Distribution parameter values}
\label{tab:dist_param}
\centering
\begin{tabular}{ccc}
\hline
Distribution & Parameter value range & Step parameter\\
\hline
Uniform & N/A & N/A \\
Gaussian & $0.5 \le \sigma \le 2.0$ & $s=0.1$ \\
Exponential & $0.5 \le \lambda \le 2.0$ & $s=0.1$ \\
Zipf & $0.5 \le \rho \le 3.0$ & $s=0.1$ \\
\hline
\end{tabular}
\end{table}

Once a tenant begins to access her data, the system starts estimating the distribution of the keys. 
A hypothesis here is that the key distribution approximates one of the probabilistic distributions we consider in this study, including uniform, Zipf, exponential, and Gaussian. 
Past studies assumed a Zipf distribution to model the workload in a cloud and caching~\cite{dan2014dynamic,Zhang2014ProactiveWM,Golrezaei:2014,GlobeTraff,YuDWC06}.
In the thorough workload study based on cache miss rate~\cite{Wires:2014}, it is observed that many workloads can be modeled using Zipf and uniform distribution.
We also take the exponential distribution into account, based on the observation 
that this distribution would be effective to approximate certain models (e.g., the reference rank model of Internet media objects)~\cite{Guo:2008}.   
Additionally, we consider Gaussian distributions for estimating the access pattern.

The estimation of distribution takes place through the KS test, which has been broadly employed for comparing empirical distributions~\cite{Choi:LBNL:2013}.
In the KS test, the resulted $p$-value indicates the similarity of the two distributions in question.
In our context, two samples in comparison are: (1) the collection of query samples for a tenant, and (2) the synthetic samples derived from the tenant's key space using a distribution model.
A higher $p$-value indicates the tenant access pattern is closer to the synthetic distribution in comparison.

Estimating the degree of uniformity is straightforward and can be done through a single KS test, since any sample set from the uniform distribution have the identical property.
In contrast, measuring the similarity against the non-uniform distributions would be  complicated and the property of the samples can be varied by the distribution-specific parameter. 
For example, a Zipf distribution becomes unique with parameter $\rho$, based on $y \sim x^{-\rho}$.
Similarly, the exponential distribution has parameter $\lambda$ from $y \sim e^{-\lambda x}$, and Gaussian is defined with a variance $\sigma$ in the standard distribution where $\mu$=0.
We consider a broad range of values for the distribution-specific parameters for greater accuracy in the estimation process, as summarized in Table~\ref{tab:dist_param}.
%
From the table, the parameter values are ranged from the min to max, which determines the skewness of the distribution.
For example, higher $\rho$ and $\lambda$ values indicate greater skewness for Zipf and exponential, respectively.
In contrast, a lower $\sigma$ implies a greater skewness for Gaussian. 


To measure the similarity against the non-uniform distributions, 
we define a step parameter $s=0.1$ as the granularity in comparison. 
For each non-uniform distribution, a KS test is arranged for a specific parameter value, from the min to max value increased by $s$.
For example, the estimation process runs 16 independent KS tests to measure the similarity against the Gaussian distribution with a $\sigma$ value from 0.5 to 2.0, incrementing it by $s$=0.1. 
%
The estimation process continues with the exponential and Zipf distributions with their distribution parameter and $s$.
Finally, a distribution with the greatest similarity is chosen as the estimated distribution to model the tenant's access pattern. 
If the KS test fails to identify a candidate distribution for the given access samples, the uniform distribution is assumed as the access pattern that requires the largest cache space compared to the other distributions as the fallback scenario.

\begin{table*}[!tb]
\caption{Impact of the number of samples to KS test}
\label{tab:kstest}
\centering
\begin{tabular}{cccccccccc}
\hline
\# samples	&	100	& 200	& 300	& 500	& 1000	& 2000	& 5000	& 7000	& 10000 \\
\hline
Correct dist.&									
		97\%	&100\%	&100\%	&100\%	&100\%	&100\%	&100\%	&100\%	&100\%  \\
Exact param. &									
		58\%	&53\%	&68\%	&66\%	&76\%	&81\%	&93\%	&93\%	&97\%  \\
$\epsilon \le 0.1$ &
		75\%	&78\%	&93\%	&90\%	&100\%	&100\%	&100\%	&100\%	&100\%  \\
$\epsilon \le 0.2$ &
		90\%	&85\%	&95\%	&98\%	&100\%	&100\%	&100\%	&100\%	&100\%  \\
\hline
\end{tabular}
\end{table*}

Intuitively, using a larger number of query samples would be helpful for estimating the access pattern more accurately. 
Table~\ref{tab:kstest} shows the impact of the number of samples in  the estimation process.
In the table, $\epsilon$ stands for the difference between the actual distribution parameter value ($p$) and the estimated one ($\hat{p}$) 
(i.e., $\epsilon = |p-\hat{p}|$).
As seen from the table, it is possible to estimate the distributions correctly only with 200 samples. 
However, 15\% of the estimations show $\epsilon > 0.2$, which may lead to a non-negligible error in the next stage for predicting cache hit rates.
It becomes quite stable with 1,000 samples or more, and 100\% of the estimates are bounded to $\epsilon \le 1.0$. 
Since we assume a high access rate (e.g., eCommerce services), collecting 1,000 samples would not be a big deal and could be made within a short time interval. 

\subsection{Cache Hit Rate Prediction}

\begin{table*}[!tb]
\caption{Learning data for cache hit rate prediction}
\label{tab:learning_data}
\centering
\begin{tabular}{|c|c|c|}
\hline
             & \multicolumn{2}{c|}{Training} \\\cline{2-3}
Distribution & data size & parameter values\\
\hline
Uniform & \{1GB, 2GB, 4GB, 8GB\} & N/A \\
\hline
Gaussian($\sigma$) & [1GB..9GB] incremented by 1GB & $\{0.5, 1.0, 1.5, 2.0\}$ \\
\hline
Exponential($\lambda$) & [1GB..9GB] incremented by 1GB & $\{0.5, 1.0, 1.5, 2.0\}$ \\
\hline
Zipf($\rho$) & [1GB..9GB] incremented by 1GB & $\{0.5, 1.0, 1.5, 2.0, 2.5, 3.0\}$\\
\hline
\end{tabular}
\end{table*}

\begin{table*}[!tb]
\caption{Testing data for cache hit rate prediction. Note that the testing data sets are {\em disjoint} from the training data sets in Table~\ref{tab:learning_data} with no overlaps. }
\label{tab:testing_data}
\centering
\begin{tabular}{|c|c|c|}
\hline
              & \multicolumn{2}{c|}{Testing} \\\cline{2-3}
Distribution & data size & parameter values \\
\hline
Uniform &  \{3GB, 6GB\} & N/A \\
\hline
Gaussian($\sigma$) & [1GB..9GB] by 1GB & $\{0.7, 1.2, 1.9\}$ \\
\hline
Exponential($\lambda$) & [1GB..9GB] by 1GB & $\{0.7, 1.2, 1.9\}$ \\
\hline
Zipf($\rho$) & [1GB..9GB] by 1GB & $\{0.7, 1.2, 1.9, 2.3, 2.6\}$ \\
\hline
\end{tabular}
\end{table*}

Once a query distribution is estimated, the next step in the process model is to predict the cache performance for the tenant in question.
Depending on the prediction result, the tenant's cache space can be adjusted
 to keep up with the data access pattern changes over time.
A high degree of accuracy for the prediction is thus essential and the tenant's QoS requirement may not be met otherwise.
%
In this section, we examine a set of regression techniques including SVR, GPR, and FCN, with the metric of MSE (Mean Squared Error) to evaluate the regression performance.  
For thorough analysis, we employ an extensive set of data sizes and distribution parameter values, summarized in Table~\ref{tab:learning_data} and Table~\ref{tab:testing_data}. 
Note that the data sets for training and testing are {\em disjoint} without any overlaps, as can be seen from the tables.
We also examine the models under the assumption of different cache sizes from 0.1 GB to 4 GB.
For simplicity, we assume four bytes for the key field and 100 KB for the value stored in the key-value repository;
for example, there exist 10,240 unique keys if the data size is 1 GB (i.e., $\frac{1GB}{100KB}$).

To measure  actual cache hit rates, we set up a cloud testbed using five servers in a cluster computing system installed with Apache Ignite as the in-memory caching infrastructure.
Three nodes are assigned for the cache service, and one node each for a client and backend database server.
The detailed information regarding our experimental settings can be found in Section~\ref{sec:ex_setting}.

\begin{figure*}[!tb]
 \centering
 \subfigure[Uniform (best: $C$=10, $\gamma$=1.0, MSE=22.03)] {
    \label{fig:svr_uni}
    \includegraphics[width=.35\textwidth]{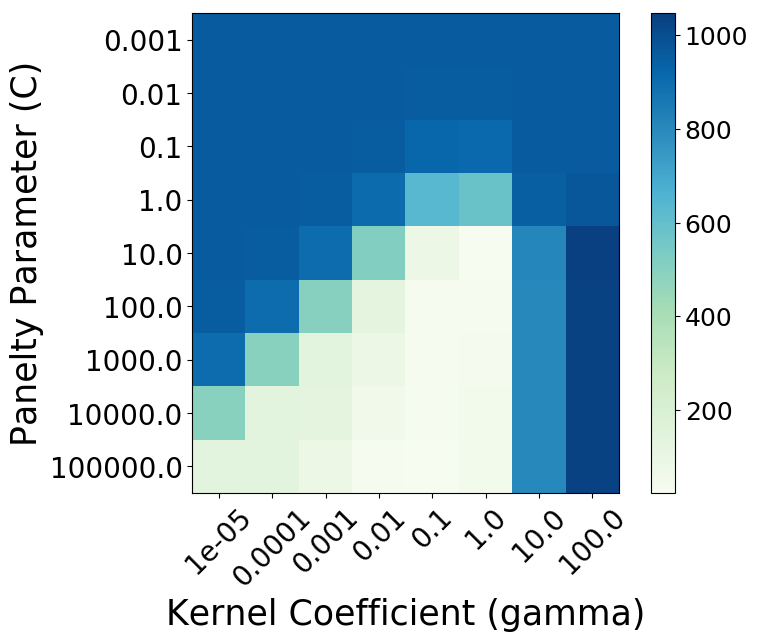}}
 \hspace{.4in}
 \subfigure[Gaussian (best: $C$=1000, $\gamma$=1.0, MSE=10.01)] {
    \label{fig:svr_gaus}
    \includegraphics[width=.35\textwidth]{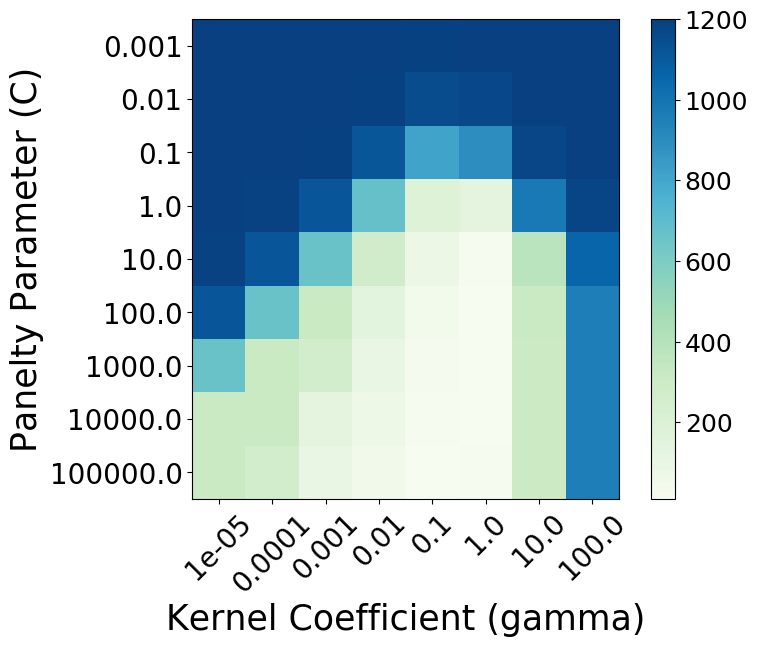}}
 \subfigure[Exponential (best: $C$=1000, $\gamma$=1.0, MSE=4.90)] {
    \label{fig:svr_exp}
    \includegraphics[width=.35\textwidth]{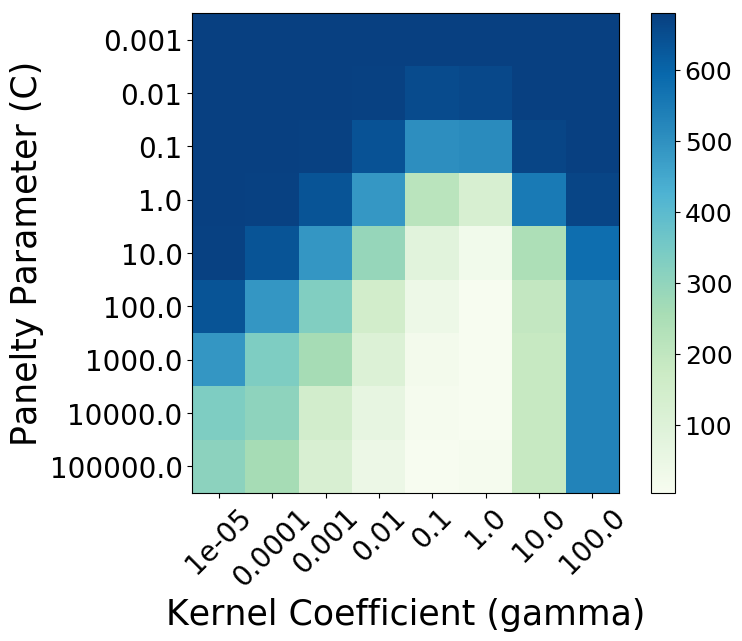}}
 \hspace{.4in}
 \subfigure[Zipf (best: $C$=10000, $\gamma$=1, MSE=1.35)] {
    \label{fig:svr_zipf}
    \includegraphics[width=.35\textwidth]{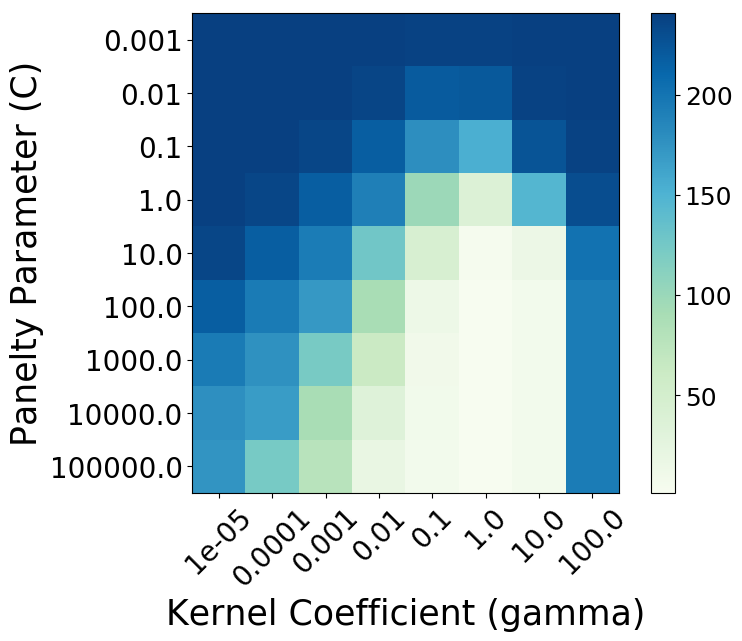}}
 \caption{SVR parameter tuning for cache hit rate regression: 
$C$=penalty parameter and $\gamma$=kernel coefficient parameter
 }
 \label{fig:svr}
\end{figure*}

\subsubsection{SVM Regression (SVR)}

SVM has been widely employed for classification and regression~\cite{Rahul14,Zhou:2013,Baek:2017}.
To predict the cache hit rate, the first regression model we examined is SVM regression (SVR).
We assumed the Radial Basis Function (RBF) kernel and the standard scaler for normalization\footnote{We  examined both standard and MinMax scalers for SVR and GPR, and observed  none of them outperforms the other completely.}.
A tricky part using SVR is to tune a set of parameters to optimize.
We examined SVR with an extensive set of the values for the penalty parameter ($C$) and the kernel coefficient parameter ($\gamma$): $10^{-3} \le C \le 10^7$ and $10^{-5} \le \gamma \le 10^3$.

Figure~\ref{fig:svr} shows the regression performance (with respect to MSE) with the combinations of the  parameter values in SVR. 
Lighter colors indicate smaller errors.
The figure shows that it needs to optimize the parameters for each distribution independently and there exists no single pair of parameters working well for all the  distributions. 
We observed that SVR works better for heavy-tail distributions (i.e., Zipf and exponential distributions) showing MSE $<$ 5.0 with the parameter tuning, but poorly works for the uniform distribution with large errors (MSE $>$ 22.0) even at best.

\begin{figure*}[!tb]
 \centering
 \subfigure[Uniform (best: $CV$=1, $LS$=10.0, MSE=18.76)] {
    \label{fig:gpr_uni}
    \includegraphics[width=.35\textwidth]{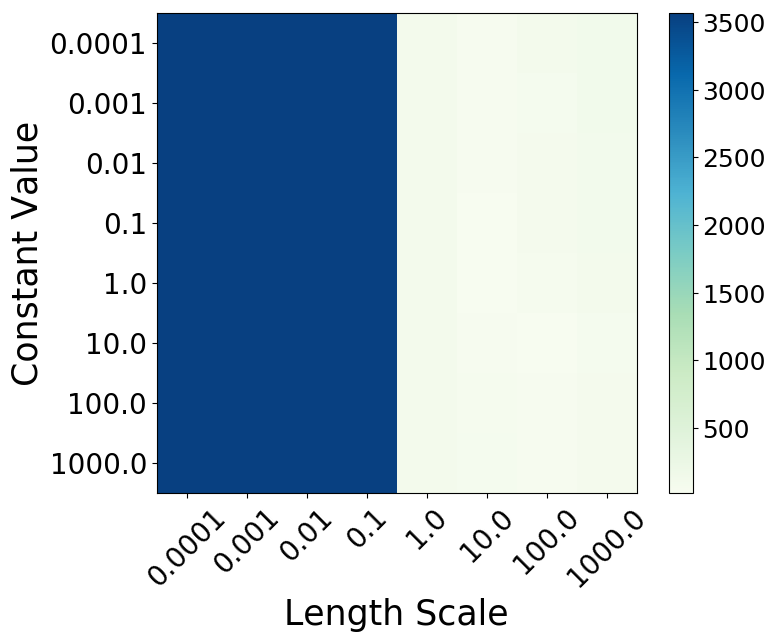}}
 \hspace{.4in}
 \subfigure[Gaussian (best: $CV$=1000, $LS$=1.0, MSE=10.90)] {
    \label{fig:gpr_gaus}
    \includegraphics[width=.35\textwidth]{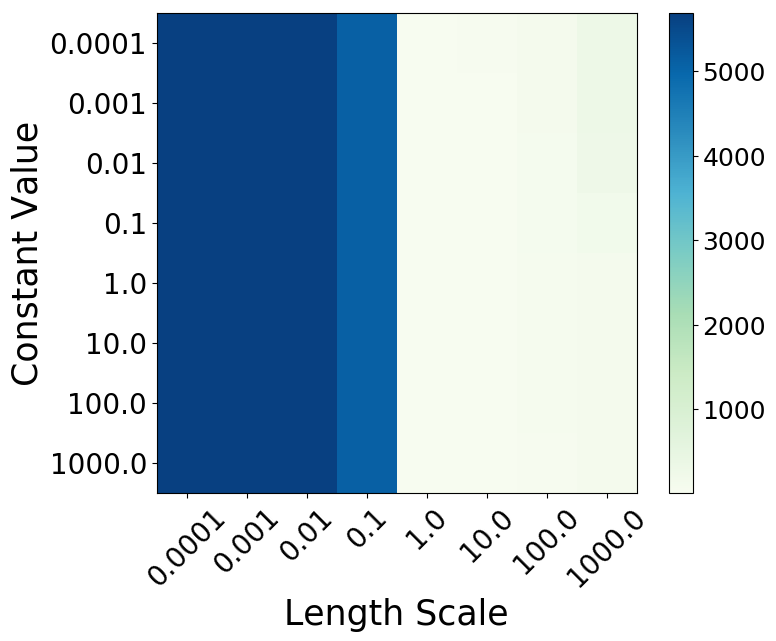}}
 \subfigure[Exponential (best: $CV$=1000, $LS$=1.0, MSE=5.3)] {
    \label{fig:gpr_exp}
    \includegraphics[width=.35\textwidth]{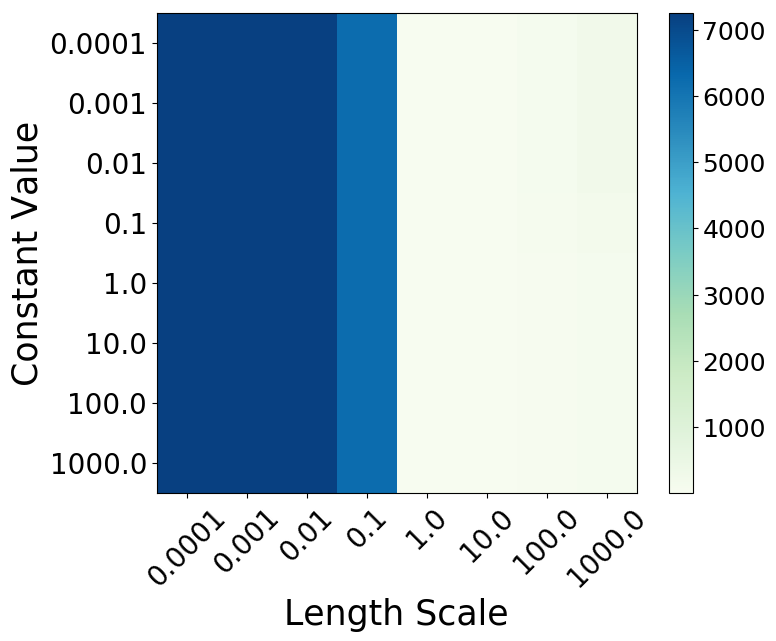}}
 \hspace{.4in}
 \subfigure[Zipf (best: $CV$=1000, $LS$=1.0, MSE=3.48)] {
    \label{fig:gpr_zipf}
    \includegraphics[width=.35\textwidth]{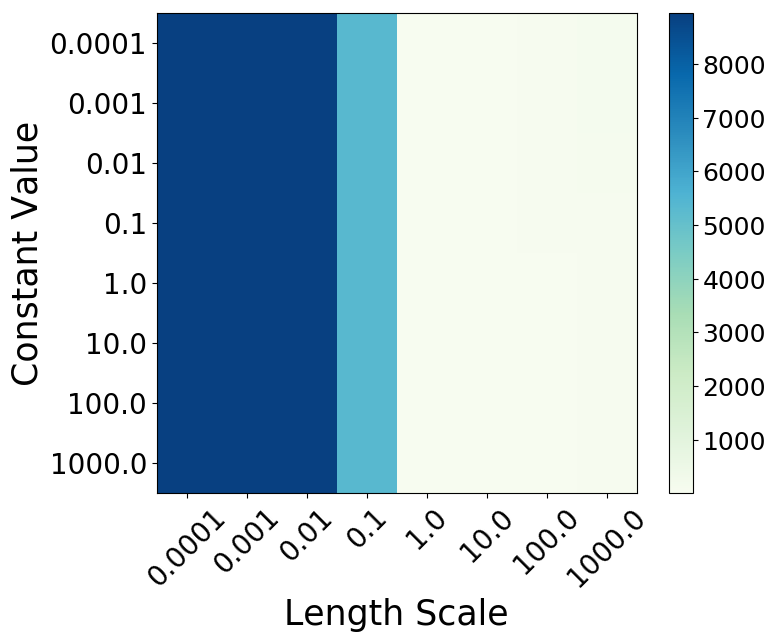}}
 \caption{GPR parameter tuning for cache hit rate regression:
	$CV$=Constant Value  and $LS$=Length Scale
 }
 \label{fig:gpr}
\end{figure*}

\subsubsection{Gaussian Process Regression (GPR)}

GPR has also been utilized for regression~\cite{gu2012spatial} and we take GPR into account to predict cache hit rates for individual distributions.
We set it up with two kernels of a constant kernel configured with Constant Value ($CV$) and a RBF kernel tuned with Length Scale ($LS$)~\cite{scikit-gpr}.
The value ranges for $CV$ and $LS$ to search the optimal are: 0.0001 $\le CV \le$ 1000 and 0.0001 $\le LS \le$ 1000.

Figure~\ref{fig:gpr} shows the regression performance when using GPR across the parameter values.
From the figure, $LS$ shows a greater impact to the regression performance, and $LS \ge$ 1.0 works much better than the other. 
The trend is similar with SVR, showing lower prediction errors for the heavily skewed distributions.
Overall, we observed no better performance from GPR compared to SVR.

\subsubsection{Regression with Fully Connected Network (FCN)}

The recent advances in deep learning has led to an introduction of many useful tools to implement learning models, promoting a wide adoption of the relevant techniques for many applications that require analyzing the data with a non-linear property.
Another regression model we investigate in this study is based on a fully connected neural network (FCN in short). 
We examined the FCN architecture with the following considerations to evaluate their impacts on the accuracy of prediction:

\begin{figure}[!tb]
\centering
\includegraphics[width=1.0\columnwidth]{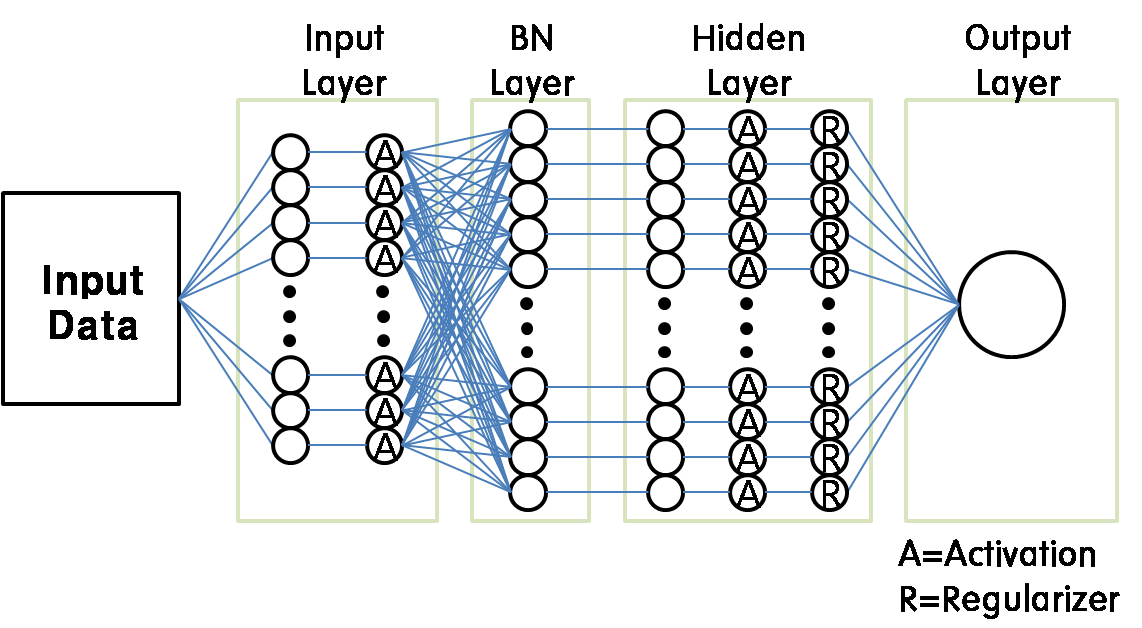}
\caption{
The FCN structure for cache hit rate prediction:
The input layer takes the input data containing three features of data size, cache size, and distribution parameter value.
We chose a single hidden layer structure based on our evaluation results.
The output is the predicted cache hit rate. 
}
\label{fig:FCN_arch}
\end{figure}

\begin{itemize}
        \item \emph{Network architecture}: To keep it simple, we designed a basic form of the FCN architecture with a single hidden layer.  The number of neurons in the hidden layer examined in the experiment is \{16, 32, 64, 128, 256\}, while we set the number of neurons in the input layer to 20.

	\item \emph{Loss functions and regularization}: Kernel regularizers play an important role to calculate penalties on layer parameters, which are then combined in the loss function to optimize the network. We evaluated L1 and L2 regularizers. We did not apply dropout and the expansion of training data for regularization.
	For the loss function, we compared Mean Absolute Error (MAE) and MSE (Mean Squared Error).

	\item \emph{Activation function}: The activation function is the non-linear transformation that determines which neurons are activated or not. We tested Sigmoid and ReLU, widely used in practice. 

	\item \emph{Number of epochs}: In an epoch, the training process performs forward and backward propagation over the entire dataset. Typically, a higher number of epochs is required for a complex dataset with a longer training time, and vice versa.
	We examined the impact of the number of epochs by setting the epoch time to one of \{500, 1000, 2000, 4000\}.

	\item \emph{Learning rate}: This rate indicates the weights update strength in back-propagation in the gradient decent. Choosing this parameter is important for training since it may not converge with a too large value, while it will be slow to converge if the learning rate is too small. 
	A typical value for learning rate is 1e-3, which also used in our experiments.

	\item \emph{Batch normalization}: The normalization layer is a key element to update the weights between neurons; that is, the weights are normalized at each update to the network.
	Our FCN model improves the regression performance with a batch normalization layer.
\end{itemize}

Figure~\ref{fig:FCN_arch} illustrates our FCN model. 
The input layer takes the input data containing three features of data size, cache size, and distribution parameter value.
As mentioned, the Batch Normalization (BN) layer is a key to optimize the overall performance and we observed a significant performance improvement with this intermediate layer. 
We evaluated a set of  structures with different numbers of hidden layers but observed insignificant  performance gaps among them; for simplicity, we chose a single hidden layer structure. 
The output is the predicted cache hit rate. 
We employed Adam as the optimizer in the implementation~\cite{Adam}. 


We conducted extensive experiments to evaluate the impact of the parameters, including the number of neurons, number of epochs, activation functions (Sigmoid and ReLU), loss functions (MAE and MSE), and regularization (L1 and L2). 
Table~\ref{tab:fcn_opt} shows the configuration performing the best for each distribution.
Overall, we observed Sigmoid and L2 regularizer work slightly better than the other options. 
MAE works better for uniform and Zipf, whereas MSE is more a suitable choice for Gaussian and exponential. 
In our experiments, the number of neurons in the hidden layer do not make considerable impacts on the regression performance.
We also observed that the epoch parameter has a slightly greater sensitivity than the number of neurons. 

\begin{table}[!tb]
\caption{Optimal configuration for  FCN model}
\label{tab:fcn_opt}
\centering
\begin{tabular}{ccccccc}
\hline
Distribution  	& \# Neuron	& Loss	& Activation 	& \# Epochs & Reg. \\
\hline
Uniform 	& 16	& MAE 	& Sigmoid 	& 4000	& L2 \\
Gaussian	& 64	& MSE	& Sigmoid	& 4000	& L2\\
Exponential  	& 64	& MSE	& Sigmoid	& 4000	& L2\\
Zipf 		& 32	& MAE 	& Sigmoid	& 500	& L2\\
\hline
\end{tabular}
\end{table}

\subsubsection{Comparison of Prediction Performance}

\begin{figure*}[!tb]
 \centering
 \subfigure[Data=3 GB] {
    \label{fig:reg_uni1}
    \includegraphics[width=.40\textwidth]{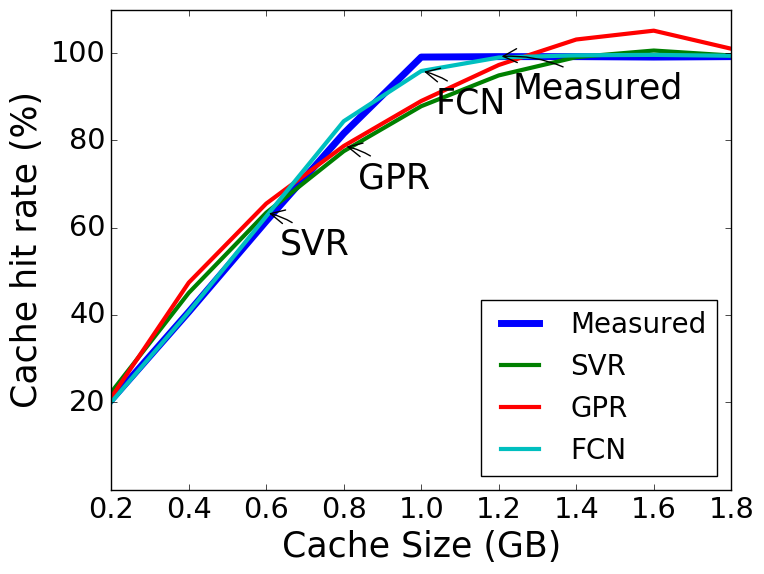}}
 \hspace{.2in}
 \subfigure[Data=6 GB] {
    \label{fig:reg_uni2}
    \includegraphics[width=.40\textwidth]{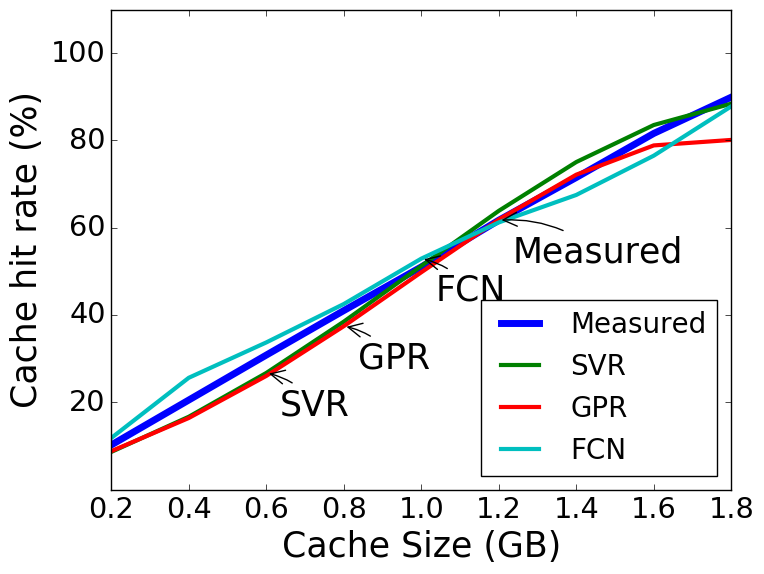}}
 \subfigure[2 tenants w/ cache=1 GB/server] {
    \label{fig:reg_uni3}
    \includegraphics[width=.40\textwidth]{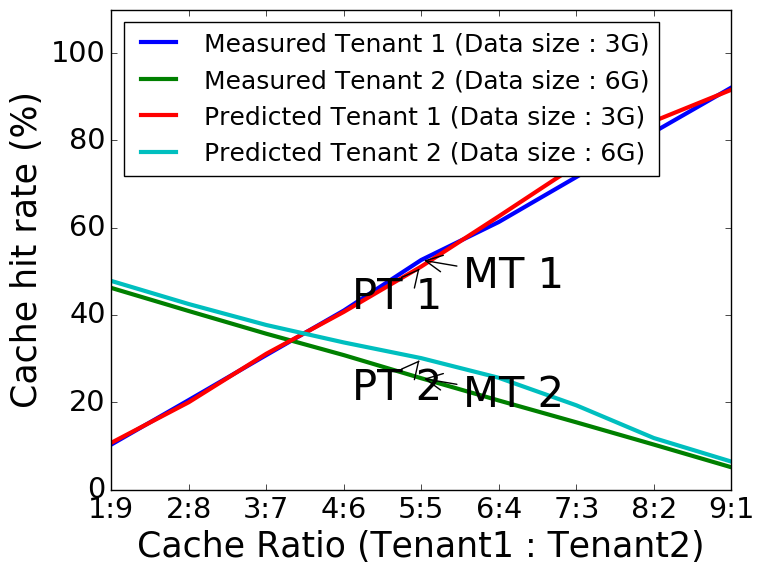}}
 \hspace{.2in}
 \subfigure[2 tenants w/ cache=2 GB/server] {
    \label{fig:reg_uni4}
    \includegraphics[width=.40\textwidth]{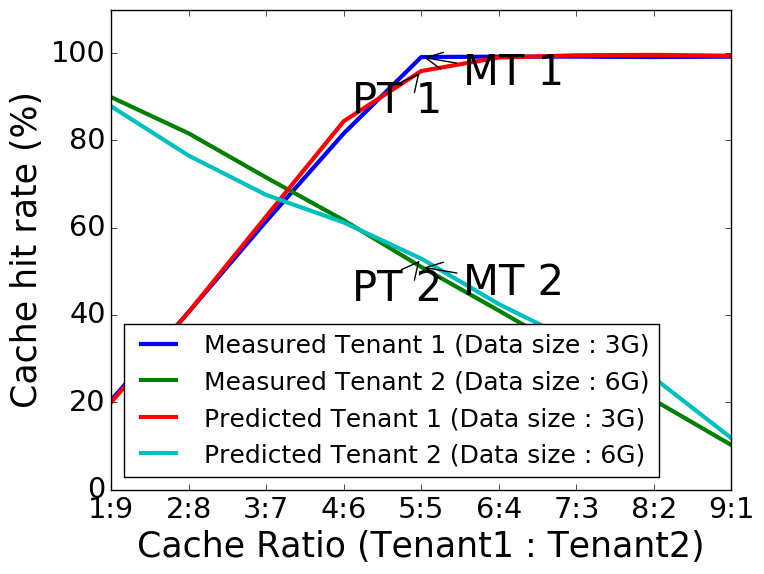}}
 \caption{Regression performance with Uniform distributions: (a) and (b) show the regression performance over different cache sizes for a tenant, and (c) and (d) show the performance for two tenants over different cache allocation ratios. 
 }
 \label{fig:reg_uni}
\end{figure*}


We next report the performance comparison for the three regression models. 
We present a subset of the experimental results due to the space reason, but the other results also show almost the same trend with insignificant differences.

Figure~\ref{fig:reg_uni} shows cache hit rates
over different cache sizes, under the assumption of the uniform access. 
Figure~\ref{fig:reg_uni1} and~\ref{fig:reg_uni2} assume 3 GB and 6 GB for the data size to be accessed, respectively. 
The $x$-axis shows the cache size per server (hence, 
the aggregated cache size is three times of the cache size per node 
as we assume three cache servers in our experiments).
While the regression models work quite well, we can see that FCN slightly outperforms the others.

Figure~\ref{fig:reg_uni3} and~\ref{fig:reg_uni4} demonstrate cache hit rates for two tenants over the different cache size configurations with respect to the cache allocation ratio in the $x$-axis.
That is, the ratio of 1:9 in $x$-axis indicates that 10\% of the cache space is allocated to one tenant and the rest of the cache space (90\% of the total space) is assigned to the other.
The regression model used in this experiment is FCN.
Figure~\ref{fig:reg_uni3} compares the measured and predicted cache hit rates when using a 1 GB  cache memory, while Figure~\ref{fig:reg_uni4} shows the result with a 2 GB cache memory, under the assumption of the uniform access.
The plots show the regression performs very well for two independent tenants.

\begin{table}[!tb]
\caption{Regression performance (MSE)}
\label{tab:reg_perf}
\centering
\begin{tabular}{ccccc}
\hline
Distribution & SVR & GPR & FCN \\
\hline
Uniform & 22.03	& 18.76	& 1.70 \\
Gaussian & 10.01	& 10.90	& 4.87 \\
Exponential & 4.90	& 5.35	& 0.66 \\
Zipf &	1.35	& 3.48	& 1.43 \\
\hline
\end{tabular}
\end{table}

We also conducted a set of experiments for the non-uniform distributions and observed similar trends.
We omit the presentation of the results due to the space reason. 
Table~\ref{tab:reg_perf} summarizes the best performance of each regression model, in which we can see that SVR and GPR work poorly for the uniform distribution.
A simple FCN model works consistently outperforming the other  techniques.
In terms of the training and prediction complexities, FCN showed greater overheads than SVR and GPR. 
The FCN training complexity is quite high (with no use of GPUs) showing two orders of magnitudes higher.
Although the prediction overhead for FCN is slightly greater than the others, we observed that a single prediction could be made within 0.26 sec on a commodity computer (equipped with Intel core i5).

\subsection{Dynamic Cache Resizing}

As a result of the cache hit rate prediction, the cache management function determines whether the cache space needs to be resized or not.
The procedure for resizing the tenant cache space is straightforward with the given distribution $d$, as follows: 

\begin{enumerate}[(i)]
	\item If $H(C_{T_k}, d) < H_{T_k} - \delta_1$, then the minimal cache space $s$ such that $H(C_{T_k}+s, d) \ge H_{T_k}$ + $\delta_1$  is additionally allocated to tenant $k$;  
	\item If $H(C_{T_k},d) > H_{T_k} + \delta_2$, then the maximal cache space $s$ such that $H(C_{T_k}-s, d) \ge H_{T_k}$ + $\delta_2$ is returned from the tenant to the system;  
	\item Otherwise, cache size for tenant $k$ remains the same.
\end{enumerate}

Here, $\delta$ parameters are configurable to consider safety margins.
The worst scenario is no more cache space available in the system in case (i); 
if this is the case, we assume that the system needs to install more resources to expand the cache space in the system to meet the SLA goals for the entire tenants. 

\section{Evaluation}
	\label{sec:eval}

We designed a set of experiments to validate the operation of the dynamic cache management with the presented estimation and prediction functions in the previous section. 
In this section, we report our evaluation results conducted on a real cluster system, with (1) the synthetic traces based on different distributions, and (2) the YCSB benchmark tool.
We first describe the experimental settings, and then discuss the experimental results with the metrics  of cache hit rate and response time to measure the performance.  

\subsection{Experimental Settings}
	\label{sec:ex_setting}

\begin{figure}[!tb]
\centering
\includegraphics[width=1.0\columnwidth]{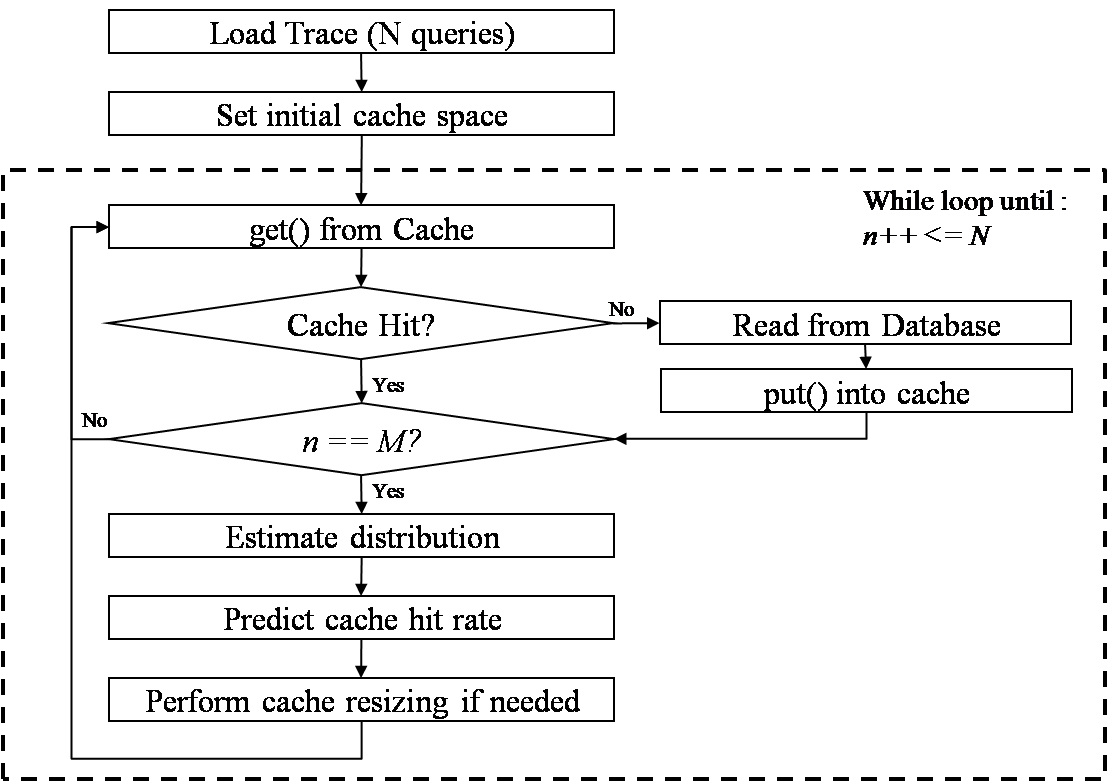}
\caption{
Flow chart of the experiment procedure
}
\label{fig:eval_flow}
\end{figure}

\begin{table}[!tb]
\caption{Experimental setting for synthetic traces. Note that the testing traces were chosen from out of the training data sets.}
\label{tab:ex_syn_conf}
\centering
\begin{tabular}{cccc}
\hline
Test case	& Distribution	& Data size	& Parameter \\
\hline
$T_1$	& Uniform	& 3 GB	& -- \\
$T_2$	& Gaussian	& 3 GB	& $\sigma=$0.7 \\
$T_3$	& Exponential	& 3 GB	& $\lambda=$0.7 \\
$T_4$	& Zipf		& 3 GB	& $\rho=$0.7 \\
\hline
\end{tabular}
\end{table}

\begin{figure*}[!tb]
 \centering
 \subfigure[Uniform ] {
    \label{fig:ex_syn_hit_uni}
    \includegraphics[width=.40\textwidth]{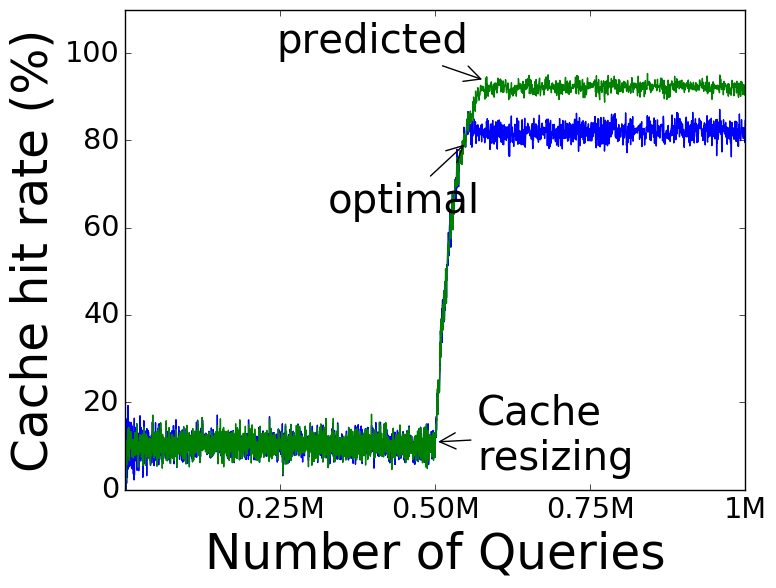}}
 \hspace{.2in}
 \subfigure[Gaussian ] {
    \label{fig:ex_syn_hit_gaus}
    \includegraphics[width=.40\textwidth]{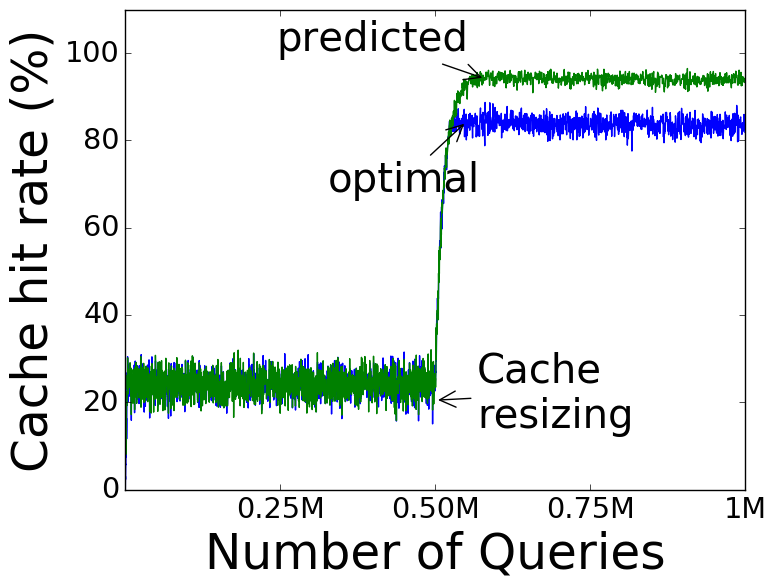}}
 \subfigure[Exponential] {
    \label{fig:ex_syn_hit_exp}
    \includegraphics[width=.40\textwidth]{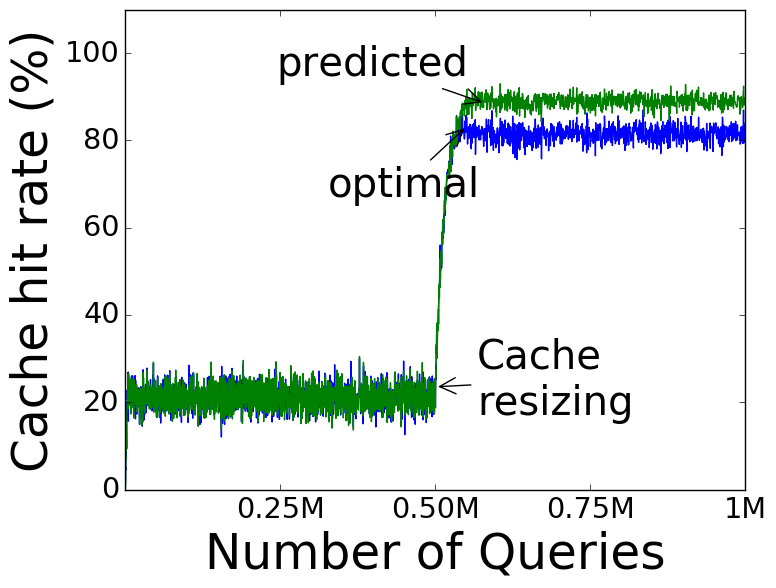}}
 \hspace{.2in}
 \subfigure[Zipf ] {
    \label{fig:ex_syn_hit_zipf}
    \includegraphics[width=.40\textwidth]{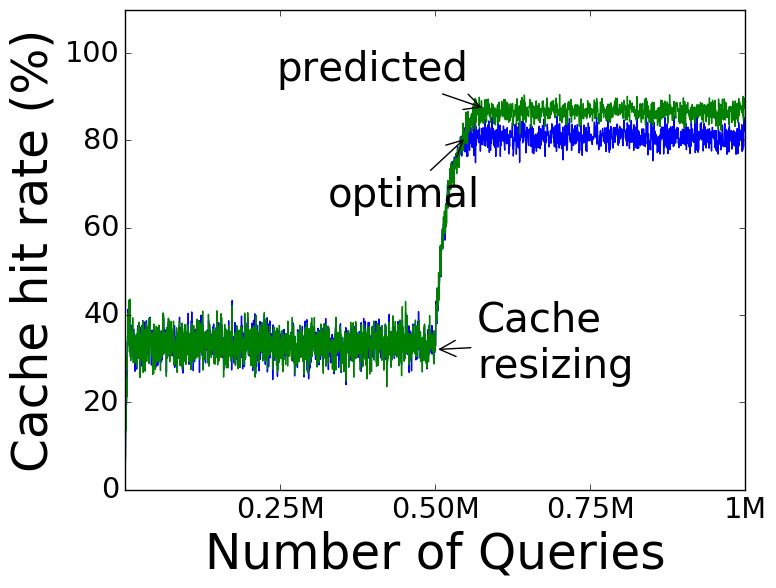}}
 \caption{Cache hit rates before and after resizing cache size with synthetic traces 
(minimum hit rate requirement=80\%, $\delta$=0.05)
 }
 \label{fig:ex_syn_hit}
\end{figure*}

\begin{figure*}[!tb]
 \centering
 \subfigure[Uniform ] {
    \label{fig:ex_syn_time_uni}
    \includegraphics[width=.40\textwidth]{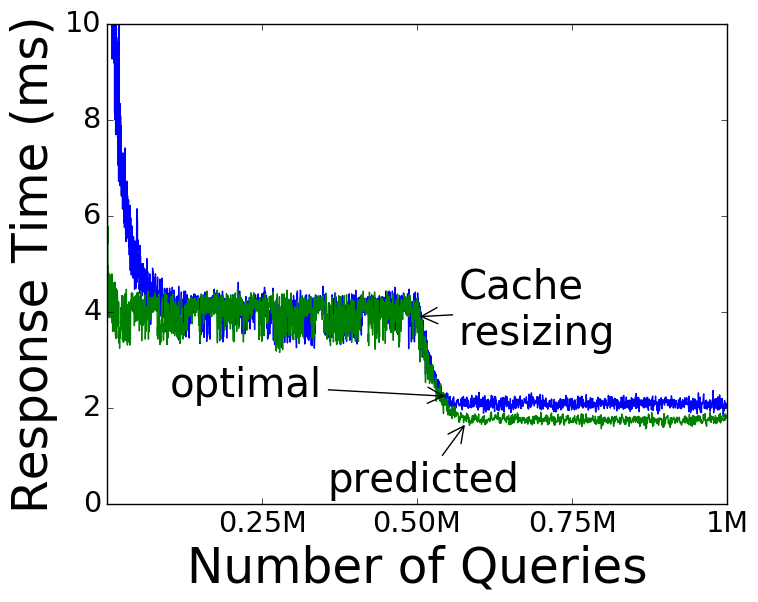}}
 \hspace{.2in}
 \subfigure[Gaussian ] {
    \label{fig:ex_syn_time_gaus}
    \includegraphics[width=.40\textwidth]{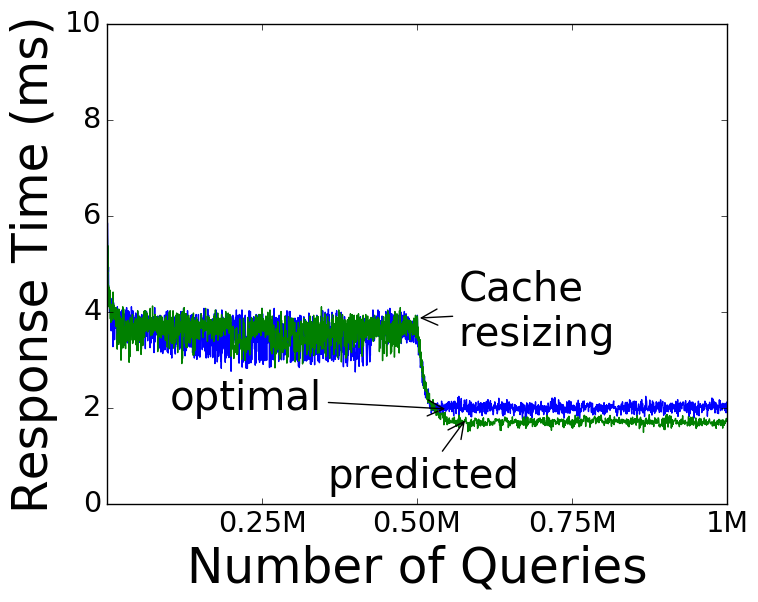}}
 \subfigure[Exponential] {
    \label{fig:ex_syn_time_exp}
    \includegraphics[width=.40\textwidth]{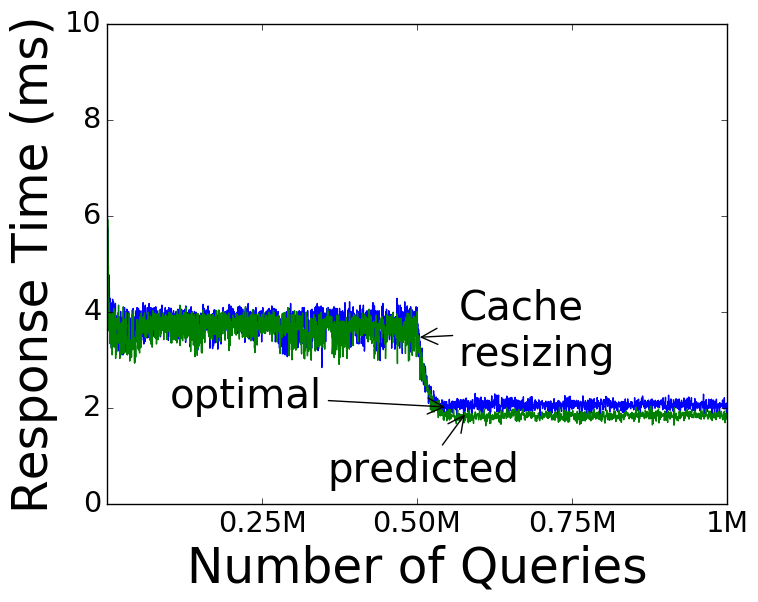}}
 \hspace{.2in}
 \subfigure[Zipf ] {
    \label{fig:ex_syn_time_zipf}
    \includegraphics[width=.40\textwidth]{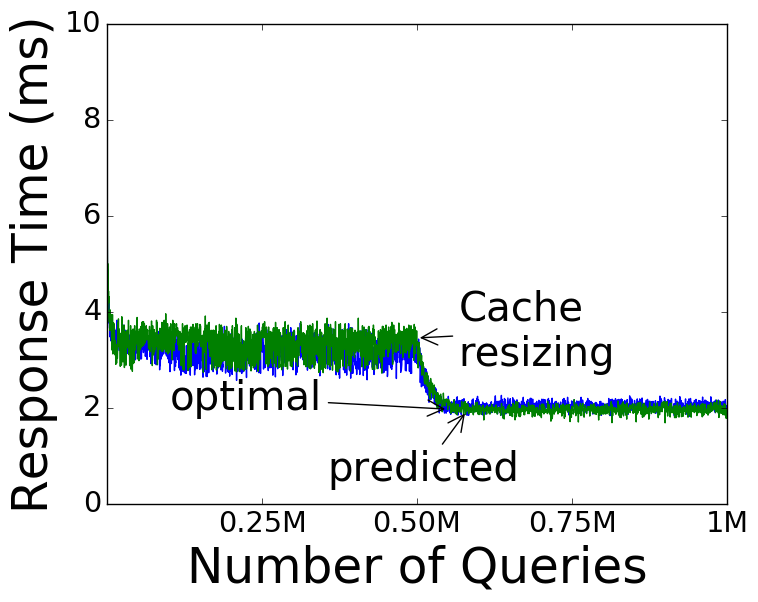}}
 \caption{Response times before and after resizing cache size with synthetic traces
(minimum hit rate requirement=80\%, $\delta$=0.05)
 }
 \label{fig:ex_syn_time}
\end{figure*}

We conducted our experiments in a computing cluster ({\tt elephant.tamuc.edu}) that consists of 27 nodes mounted in a rack. Each node consists of 4 CPU cores, 8 GB memory, and 2 TB hard disk storage.
The nodes are interconnected via a Gigabit Ethernet switch. 
We installed Apache Ignite 1.8.0~\cite{ApacheIgnite} and MariaDB~\cite{MariaDB} as the in-memory cache infrastructure and the backend DBMS, respectively. 
We configured three nodes for the in-memory cache service, each of which is configured with 6 GB cache space, and hence, the total cache space is 18 GB in the system. 
Apache Ignite provides a set of built-in eviction methods including FCFS and LRU, and we simply chose LRU for cache replacement.
One node is dedicated as a database server with MariaDB.

Figure~\ref{fig:eval_flow} illustrates the  procedure for experiments, the main objective of which is to see if the proposed dynamic cache management is effective to meet the tenant's QoS requirement based on the data access pattern and the specified cache hit rate requirement. 
We assume that the data size is 3 GB for the tenant and the required cache hit rate is 80\% at minimum.
The initial cache size is 0.1 GB/node (i.e., 0.3 GB/system with three cache servers).
Each experiment consists of $N$ queries with a certain distribution (to generate keys), and $N$=1 million by default.
For each query, the \texttt{get(key)} operation is invoked to look up the cache.
In case of cache miss, the backend server is accessed to retrieve the entry associated with the given key from the database, and the \texttt{put(key,val)} operation is executed to add a new entry to the cache.  
Once $M$ queries are serviced, the system performs the estimation and prediction functions to see if resizing of cache space is needed.
We set $M=\frac{N}{2}$ in our experiment to compare the performance before and after the event of cache resizing.  
Since the initial cache size is too small (0.3 GB) compared to the data size (3 GB), cache resizing will be triggered to meet the desired performance requirement.
Note that the distribution estimation is performed with the KS-test and the prediction takes place using the FCN model, described in the previous section.
The parameter values used for FCN can be found from Table~\ref{tab:fcn_opt}.

\subsection{Experiments with Synthetic Traces}
	\label{sec:ex_syn}

We first present the experimental results conducted with a set of synthetic traces with different distribution models.
For the experiments, we installed Yardstick-ignite as a benchmark tool.
Yardstick-Ignite provides 8 types of benchmark tests, and  
we utilized {\tt PutGetTxBenchmark}  providing transactional distributed cache put and get operations.
Table~\ref{tab:ex_syn_conf} shows the test cases prepared for the experiments with the synthetic data. 
As noted earlier, the data sets used for training and testing are {\em disjoint} without any overlaps, as summarized in Table~\ref{tab:learning_data} and Table~\ref{tab:testing_data}.
Thus, the data size (3 GB) and the distribution parameter values used for testing were chosen from outside the training data sets. 

Figure~\ref{fig:ex_syn_hit} 
shows the cache hit rates before and after the cache resizing. 
In the figure, ``optimal'' resizes the cache based on the measurement data without relying on the prediction, while ``predicted'' manages the cache based on our prediction procedure.
For the prediction, we set the safety margin to 5\% (i.e., $\delta$=0.05) based on the observation that the max difference between the measured and predicted hit rates is less than 0.05. 
With the safety margin, our prediction-based cache management will try to adjust the cache size to make the predicted hit rate to be (80 + $\delta$)\% at minimum. 

From the figure, the initial hit rates are very low. 
Based on the estimation of the access distribution and the prediction of the hit rate, the cache size is adjusted at $n$=500,000 queries, and the cache hit rates jump up to over 80\% on average.
The figure shows that the prediction works very well and allocates a slightly greater space for the cache compared to optimal (0.1 GB greater on average), due to the safety margin.

Figure~\ref{fig:ex_syn_time} shows the corresponding response time over the number of queries.
In the initial warm up phase, the response time is very high since there is no entry in cache and every operation triggers the database access.
As soon as the entire cache space is filled in, the average response time becomes stable.
We can see that the response time significantly goes down after the cache is resized at $n$=500,000.
The response time based on the prediction is slightly lower than optimal, since the prediction-based resizing allocates the more space to cache (and hence, with greater hit rates).

\begin{table*}[!tb]
\caption{Experimental result with synthetic traces before resizing cache memory}
\label{tab:ex_syn_result_before}
\centering
\begin{tabular}{cccc}
\hline
Test	& Initial 	& Initial	& Initial  \\
case	& cache size 	& hit rate	& resp. time \\
\hline
$T_1$	& 0.1 GB	& 10.2\%	& 4.58 ms \\
$T_2$	& 0.1 GB	& 24.1\%	& 3.59 ms \\
$T_3$	& 0.1 GB	& 21.1\%	& 3.67 ms \\
$T_4$	& 0.1 GB	& 33.2\%	& 3.37 ms \\
\hline
\end{tabular}
\end{table*}

\begin{table*}[!tb]
\caption{Experimental result with synthetic traces after resizing cache memory}
\label{tab:ex_syn_result_after}
\centering
\small
\begin{tabular}{ccccccc}
\hline
Test	& Optimal & Optimal & Estimated	& Predicted	& Measured	& Measured \\
case	& cache size & hit rate & distribution	& cache size	& hit rate	& resp. time \\
\hline
$T_1$	& 0.8 GB & 81.8\% & Uniform	& 0.9 GB	& 91.8\% & 1.77 ms \\
$T_2$	& 0.4 GB & 83.7\% & Gaussian(0.7)	& 0.5 GB	& 94.0\% & 1.78 ms \\
$T_3$	& 0.5 GB & 80.6\% & Exponential(0.7)	& 0.6 GB	& 88.7\% & 1.84 ms \\
$T_4$	& 0.6 GB & 80.7\% & Zipf(0.7)	& 0.7 GB	& 86.4\% & 1.96 ms \\
\hline
\end{tabular}
\end{table*}

Table~\ref{tab:ex_syn_result_before} and Table~\ref{tab:ex_syn_result_after} summarize the experimental results with the four synthetic data sets.
We can see that our estimation process is able to identify the exact distribution and the associated parameter value. 
In the table, the measured hit rate and response time stand for the cache hit rate and response time after resizing the cache space based on the predicted cache size information, which meet the required hit rates. 

\begin{figure}[!tb]
 \centering
 \subfigure[Cache hit rate] {
    \label{fig:ex_syn_multi_hit}
    \includegraphics[width=1.0\columnwidth]{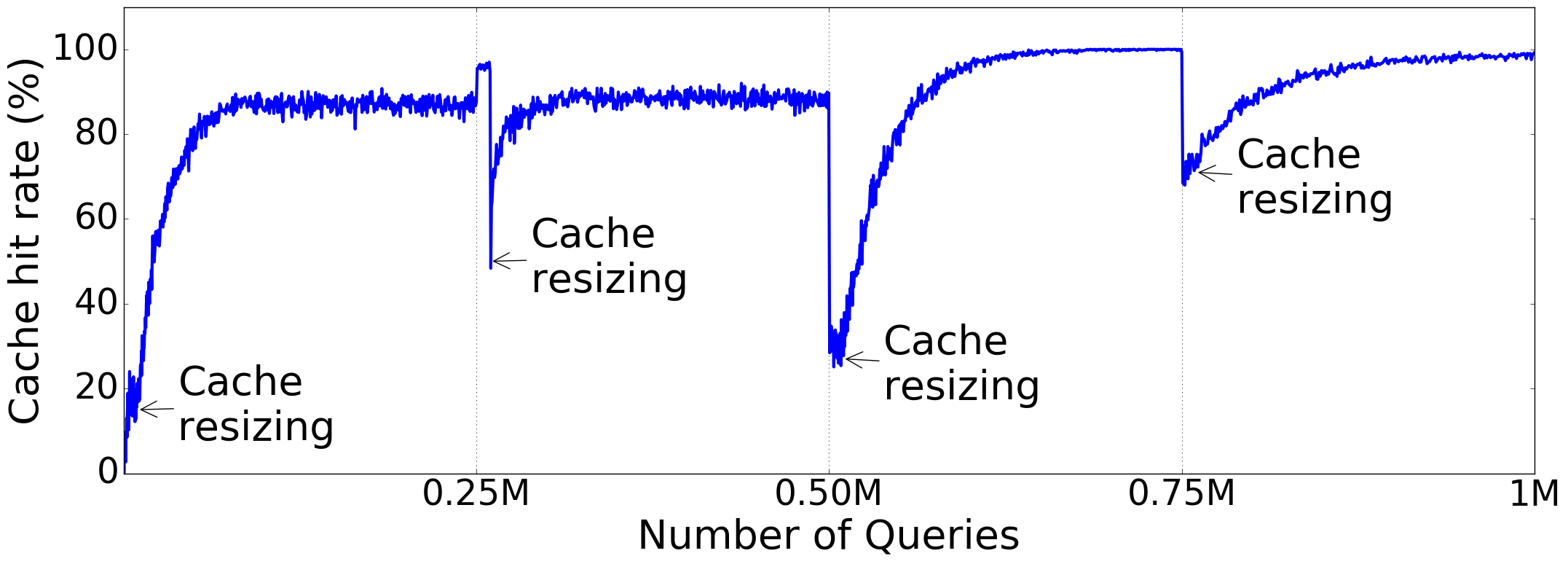}}
 \subfigure[Response time] {
    \label{fig:ex_syn_multi_time}
    \includegraphics[width=1.0\columnwidth]{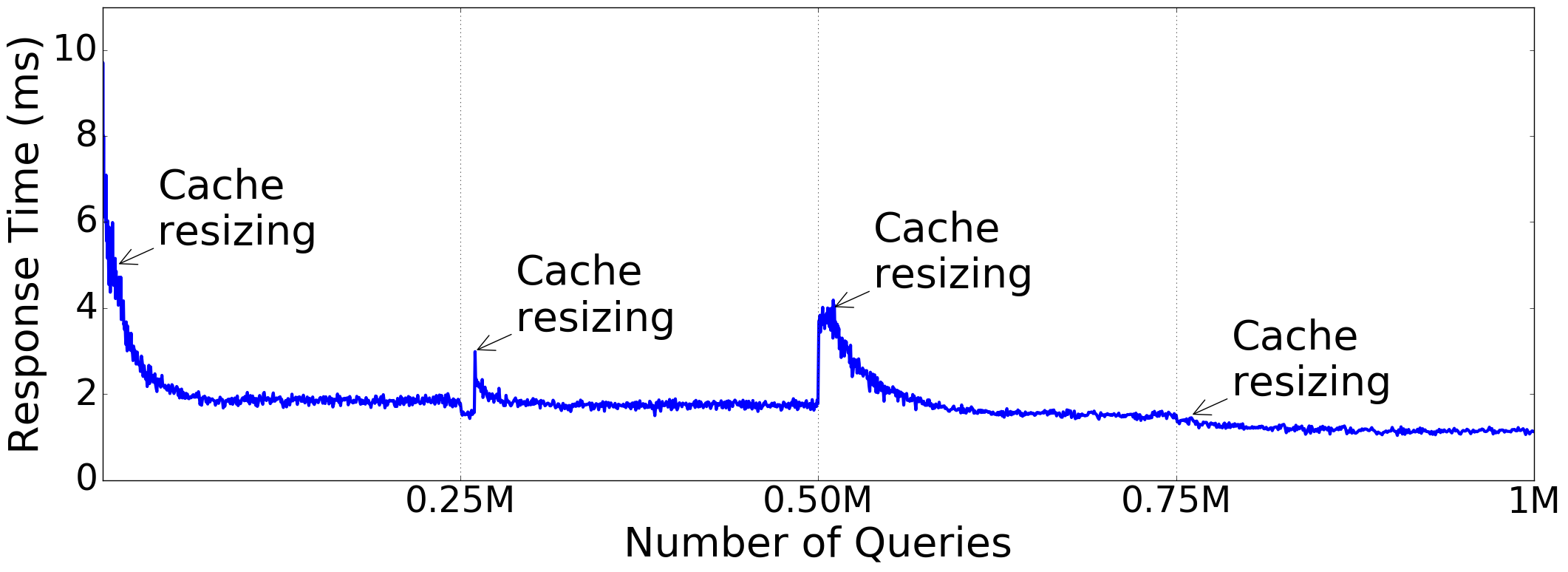}}
 \caption{Cache hit rate and response time over data access pattern changes:
The injected trace includes four different distributions in order: (1) exponential ($\lambda$=0.9), (2) Zipf ($\rho$=1.1), (3) uniform, and  (4) Gaussian ($\sigma$=1.3).
The initial cache size is 0.1 GB, the required hit rate is 80\%, and $\delta$=0.
 }
 \label{fig:ex_syn_multi}
\end{figure}

We next assume consecutive changes of the data access patterns over time. 
In this experiment, the estimation process takes place at every 10,000 queries.
If any pattern change is identified, the cache prediction process is activated and the cache space is reallocated if needed.
The injected trace includes four different distributions in order: (1) exponential ($\lambda$=0.9), (2) Zipf ($\rho$=1.1), (3) uniform, and  (4) Gaussian ($\sigma$=1.3).
The initial cache size allocated is 0.1 GB, and the required hit rate is 80\%.
We simply set $\delta$=0.0 (i.e., no safety margin) in this experiment.
The following lists the the events and observations over the temporal pattern changes: 

\begin{enumerate}[(i)]
   \item At $x$=0, the initial pattern follows exponential ($\lambda$=0.9); 
   \item At $x$=10,000, the cache is resized to 0.7 GB (from 0.1 GB); 
   \item At $x$=250,000, the pattern is changed to Zipf ($\rho$=1.1);
   \item At $x$=260,000, the cache size is reduced to 0.3 GB; 
   \item At $x$=500,000, the pattern is changed to uniform; 
   \item At $x$=510,000, the cache size increases  to 1.4 GB; 
   \item At $x$=750,000, the pattern is changed to Gaussian ($\sigma=$1.3);
   \item At $x$=760,000, the cache size is changed to 1.1 GB. 
\end{enumerate}

Figure~\ref{fig:ex_syn_multi} demonstrates the dynamic cache management over time. 
Initially, the system identifies the access pattern and resizes the cache space to 0.7 GB to meet the QoS requirement. 
Although the cache space is adjusted at $x$=10,000, it takes a time to warm up the cache as shown in the figure. 
At every time that the access pattern is changed, we can see the degradation of the cache hit rate, but the performance is restored by dynamically resizing the cache space.
At $x$=260,000 the cache size becomes shrunk (from 0.7 GB to 0.3 GB), but we can see that the cache hit rate meets the requirement.
Figure~\ref{fig:ex_syn_multi_time} shows the corresponding response time over the temporal changes.

\subsection{Experiments with YCSB}
	\label{sec:ex_ycsb}

\begin{figure}[!tb]
 \centering
 \subfigure[Cache hit rate] {
    \label{fig:ex_ycsb_hit}
    \includegraphics[width=.40\textwidth]{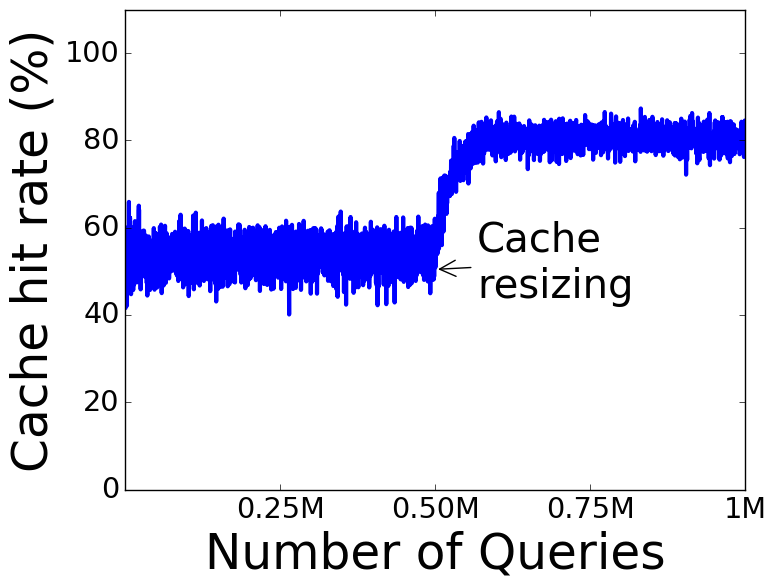}}
 \hspace{.2in}
 \subfigure[Response time] {
    \label{fig:ex_ycsb_time}
    \includegraphics[width=.40\textwidth]{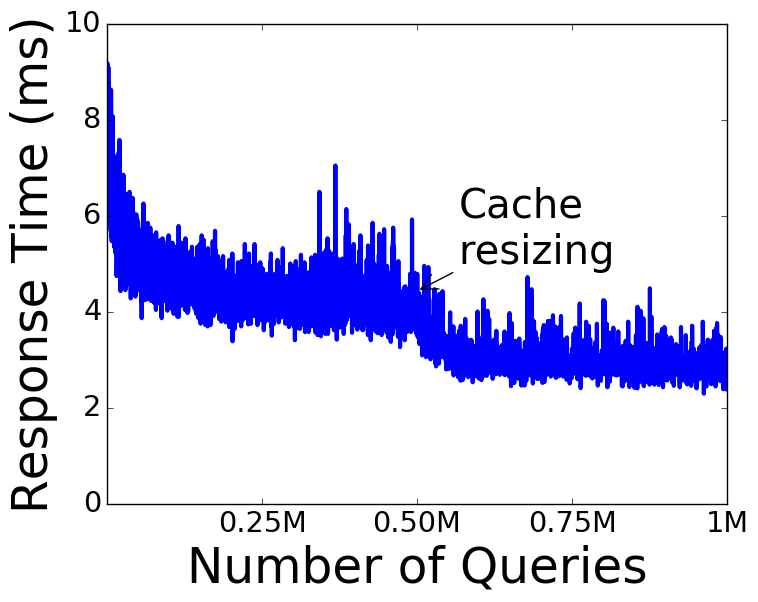}}
 \caption{Cache hit rate and response time before/after resizing cache size using YCSB benchmark (minimum hit rate requirement=80\% and $\delta$=0)
 }
 \label{fig:ex_ycsb}
\end{figure}

We next report the experimental results conducted with YCSB, a benchmark tool widely used for evaluating the performance for RDBMS and NoSQL~\cite{YCSB,FairRide}. 
We slightly modified this benchmark tool to perform the experiments as specified in the experimental procedure in Figure~\ref{fig:eval_flow}.

Figure~\ref{fig:ex_ycsb} demonstrates the cache hit rate and response time experimented with the YCSB tool.
As the experiments with the synthetic traces, we initially allocate 0.1 GB/node for the cache space, and the estimation and prediction take place at $n$=500,000. 
With the initial cache space, the cumulative cache hit rate is 53.7\% and the average response time is 4.46 msec. 
With the estimation process,  the distribution is identified as Zipf with $\rho$=1.0.
As a result of the prediction through FCN, the cache  is resized to 0.8 GB/node.
The observed cache hit rate is 80.9\% with 2.10 msec of average response time after resizing.

\section{Conclusions}
	\label{sec:conc}

The in-memory cache has been widely employed to improve data access performance in a cloud. 
Despite its  importance, the past studies were largely limited 
with the lack of well-defined models for the estimation of data access patterns and the prediction of cache performance 
for the access pattern in question.
In this paper, we proposed a learning-based approach to dynamic caching in a cloud to meet the per-tenant QoS requirement.
%
We first presented an estimation method that approximates the data access pattern to one of
four distribution models of uniform, Gaussian, exponential, and Zipf, based on the KS test.
We observed that
our estimation method works well with a high degree of accuracy even with a small number of query samples ($\ge$ 200 samples), which should be beneficial for responding to the temporal pattern changes in a timely manner.
We next presented the evaluation results of a set of regression methods including SVR, GPR, and FCN, to predict the cache hit rate based on the estimated access pattern. 
From the experiments, we observed that the FCN model outperforms the others across the  distributions. 
Finally, we evaluated our dynamic cache management method with an extensive set of synthetic traces and the YCSB benchmark.
The evaluation results show that the proposed method consistently optimizes the cache space, while preserving the tenant's QoS requirement. 

The cloud cache management consists of a set of functions 
and this work focused on the problem of cache space optimization.
Another important function in the cache management is cache eviction that has a significant impact on the data access performance. 
In this work, we simply assumed LRU as the eviction policy but the estimated access pattern would be the helpful information to improve the hit rate.
A planned future work is to investigate adaptive methods for cache eviction over access pattern changes.


\let\OLDthebibliography\thebibliography
\renewcommand\thebibliography[1]{
  \OLDthebibliography{#1}
  \setlength{\parskip}{0pt}
  \setlength{\itemsep}{2pt plus 0.3ex}
}

\bibliographystyle{unsrt}
\bibliography{paper}

\end{document}